\documentclass[
  a4paper,%
  11pt,%
  extramargin,
]{tubsartcl}

\usepackage{algorithm}
\usepackage{algpseudocode}
\usepackage{amsmath}
\usepackage{amssymb}
\usepackage{anyfontsize}
\usepackage{booktabs}
\usepackage{balance}
\usepackage{dcolumn}
\usepackage{graphicx}
\usepackage[utf8]{inputenc}
\usepackage{lmodern}
\usepackage{listings}
\usepackage{longfbox}
\usepackage{microtype}
\usepackage{multirow}
\usepackage[super]{nth}
\usepackage[numbers,sort,compress,sectionbib]{natbib}
\usepackage{pifont}
\usepackage{siunitx}
\usepackage[caption=false]{subfig}
\usepackage{tabularx}
\usepackage{tikz}
\usetikzlibrary{matrix}
\usepackage{xspace}
\usepackage[breaklinks,hidelinks,bookmarksdepth=1]{hyperref}

\newcommand{\clrLight}{tuBlueMedium20}
\newcommand{\clrMedium}{tuBlueMedium60}
\newcommand{\clrDark}{tuBlueMedium100}

\addtokomafont{disposition}{\color{\clrMedium}}

\newcommand{\clrImage}{images/title}
\newcommand{\reportno}{sec-2019-01}

\newcommand{\btw}{German federal election\xspace}

\lstdefinestyle{tweet}{
  belowcaptionskip=1\baselineskip,
  frame=single,
  xleftmargin=\parindent,
  basicstyle=\scriptsize\ttfamily,
  breaklines=true,
  breakautoindent=false,
  literate=%
    {Ö}{{\"O}}1
    {Ä}{{\"A}}1
    {Ü}{{\"U}}1
    {ß}{{\ss}}1
    {ü}{{\"u}}1
    {ä}{{\"a}}1
    {ö}{{\"o}}1
    {…}{\dots}1
    {~}{{\textasciitilde}}1
}

\newcommand{\rot}{\rotatebox{90}}

\newcommand{\perc}[1]{\SI{#1}{\percent}}
\newcommand{\code}[1]{{\footnotesize\ttfamily #1}}

\newcommand{\cross}{\ding{61}\xspace}

\newcommand{\hashtag}[1]{\emph{\small \##1\xspace}}
\newcommand{\mention}[1]{\emph{\small @#1\xspace}}
\newcommand{\party}[1]{\emph{\@#1\xspace}}

\newcommand{\alg}[1]{\hyperref[#1]{Algorithm~\ref{#1}}}
\newcommand{\tab}[1]{\hyperref[#1]{Table~\ref{#1}}}
\newcommand{\fig}[1]{\hyperref[#1]{Figure~\ref{#1}}}
\newcommand{\figs}[1]{\hyperref[#1]{Figures~\ref{#1}}}
\newcommand{\sect}[1]{\hyperref[#1]{Section~\ref{#1}}}
\newcommand{\sects}[1]{\hyperref[#1]{Section~\ref{#1}}}
\newcommand{\appx}[1]{\hyperref[#1]{Appendix~\ref{#1}}}

\newcommand{\eg}{e.g.,\xspace} %
\newcommand{\ie}{i.e.,\xspace} %

\definecolor{tubsRed}{cmyk}{0.1,1.0,0.8,0.0}

\newcommand{\tablesize}{\small}

\newcommand{\tagline}{Political Elections Under (Social) Fire?}
\newcommand{\ourtitle}{\tagline\\
                       Analysis and Detection of Propaganda on Twitter}

\newcommand{\ourkeywords}{%
Twitter, Trolls, Bots, Propaganda, German Federal Election}

\newcommand{\ansgar}{Ansgar Kellner\xspace}
\newcommand{\lisa}{Lisa Rangosch\xspace}
\newcommand{\chris}{Christian Wressnegger\xspace}
\newcommand{\knrd}{Konrad Rieck\xspace}

\newcommand{\authors}{\ansgar, \lisa, \chris, and \knrd}

\AtBeginDocument{
    \hypersetup{
        pdfinfo={
            Author={\authors},
            Subject={Propaganda on Twitter},
            Title={\ourtitle},
            Keywords={\ourkeywords},
            Creator={},
            Producer={},
        },
        pdfstartview=Fit,
        pdfpagemode=UseOutlines,
        bookmarksopen=true
    }
}

\newcommand{\timespan}{January to September 2017\xspace}
\newcommand{\numdays}{\num{268}\xspace}

\newcommand{\approxtotaltweets}{\num{9.5}~million\xspace}

\newcommand{\numtweetswithoutfdp}{\num{8845879}\xspace}
\newcommand{\perctweetswithoutfdplost}{\perc{7.56}\xspace}

\newcommand{\numinterestingusers}{\num{23949}\xspace}
\newcommand{\numminimumnumberoftweets}{\num{30}\xspace}

\newcommand{\numtopimagefirst}{\num{6027}\xspace}
\newcommand{\numtopimagesecond}{\num{2911}\xspace}
\newcommand{\numtopimagethird}{\num{1844}\xspace}
\newcommand{\numtopimagefourth}{\num{1680}\xspace}
\newcommand{\numtopimagefifth}{\num{1469}\xspace}

\newcommand{\numtrollsinlist}{\num{2752}\xspace}
\newcommand{\numtrollsfound}{\num{79}\xspace}
\newcommand{\perctrollsfound}{\perc{0.02}\xspace}

\newcommand{\numtrolltweets}{\num{9309}\xspace}
\newcommand{\perctrolltweets}{\perc{0.06}\xspace}
\newcommand{\perctrolltweetssmallerthanthirtyfive}{\perc{79.75}\xspace}
\newcommand{\numtrolltweetsorig}{\num{3520}\xspace}

\newcommand{\numtrolltweetretweet}{\num{4341}\xspace}
\newcommand{\numtrolltweetsquote}{\num{1448}\xspace}

\newcommand{\numtrollretweets}{\num{4341}\xspace}
\newcommand{\numtrollretweetsbytrolls}{\num{16}\xspace}

\newcommand{\numtrollquotes}{\num{1448}\xspace}

\newcommand{\numfeatures}{\num{44}\xspace}
\newcommand{\numcrossvalrepetitions}{\num{100}\xspace}

\newcommand{\numtraininghumans}{\num{874}\xspace}
\newcommand{\numtrainingbots}{\num{505}\xspace}

\newcommand{\numpredicthumans}{\num{20157}\xspace}
\newcommand{\numpredictbots}{\num{2414}\xspace}

\newcommand{\numtotalhumans}{\num{21030}\xspace}
\newcommand{\numtotalbots}{\num{2919}\xspace}

\newcommand{\perctotalhumans}{\perc{87.81}\xspace}
\newcommand{\perctotalbots}{\perc{12.19}\xspace}

\begin{document}
    \begin{titlepage}
        \showtubslogo
        \showlogo{\includegraphics[width=5.6cm]{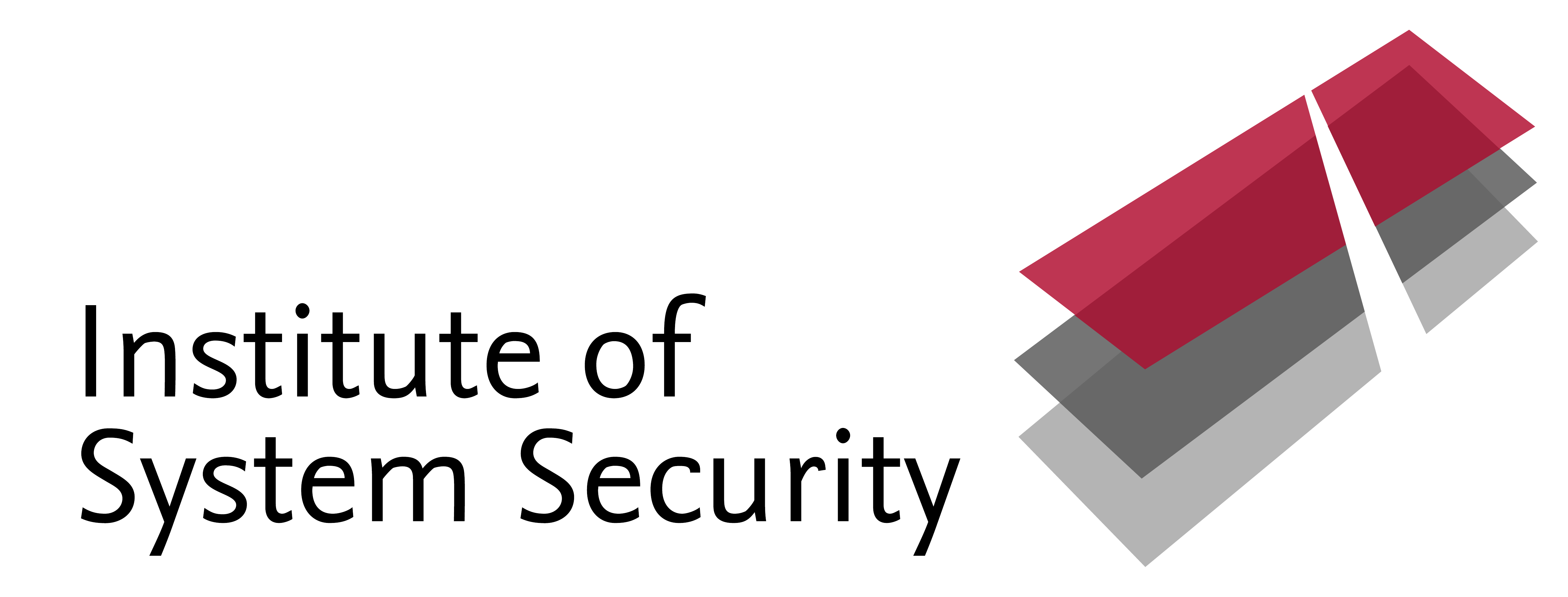}}

        \begin{titlerow}[bgimage=\clrImage]{2}
        \end{titlerow}

        \begin{titlerow}[bgcolor=\clrDark, fgcolor=tubsWhite]{1}
            \raggedright \noindent \fontsize{23}{28}\selectfont
            \ourtitle
        \end{titlerow}

        \begin{titlerow}[bgcolor=\clrLight]{4}
            \raggedright \noindent \fontsize{15}{20}\selectfont
            \authors\\
            \vspace{6mm}
            Computer Science Report No. \reportno\\
            Technische Universität Braunschweig\\
            Institute of System Security
        \end{titlerow}
    \end{titlepage}

    \newpage
    ~
    \vfill \noindent {\sffamily
    Technische Universität Braunschweig\\
    Institute of System Security\\
    Rebenring 56\\
    38106 Braunschweig, Germany}

    \newpage

    \null\vspace{-5mm}
    \begin{abstract}
        \subsection*{Abstract}
        For many, social networks have become the primary source of news,
        although the correctness of the provided information and its
        trustworthiness are often unclear. The investigations of the 2016 US
        presidential elections have brought the existence of external
        campaigns to light aiming at affecting the general political public
        opinion.
        In this paper, we investigate whether a similar influence on
        political elections can be observed in Europe as well.
        To this end, we use the past German federal election as an indicator
        and inspect the propaganda on Twitter, based on data from a period
        of
        \numdays~days.
        We find that \numtrollsfound~trolls from the US~campaign have also
        acted upon the \btw spreading right-wing views.
        Moreover, we develop a detector for finding automated behavior that
        enables us to identify \numpredictbots~previously unknown~bots.
    \end{abstract}

    \clearpage

\section{Introduction}
\label{sec:introduction}

The use of social media for propaganda purposes has become an integral part
of cyber warfare~\citep{aro16}. Most prominently, in 2016 the
US~presidential elections have been targeted by a Russian interference
campaign on Twitter~\citep{BadFerLer18}.
However, the use of online propaganda is not an isolated phenomenon, but
a global challenge~\citep{ShiJiaDri17, RamFerPin18, StiBleLieStr18}. The
effect of political propaganda and fake news is further amplified by
journalists that use Twitter to acquire ``cutting-edge information''
when chasing down trending topics for their next story~\cite{BroGra12,
BovMak19}, and distribute them via traditional~media.

In this paper, we investigate whether a similar influence on political
elections can be observed in Europe as well and thus analyze the Twitter
coverage of the \btw~(\emph{Bundestagswahl}) to figure out if the public
opinion has been influenced and by how much.
To this end, we have collected \approxtotaltweets tweets related to the
hashtags of all major German parties over \numdays~days, from \timespan.
In contrast to earlier work on the influence on
Twitter~\citep{YeWu10, BakHofMasWat11, RiqGon16}, we focus on basic
features
that can directly be derived from the Tweets and their metadata, such as the
number of retweets or quotes.
The mere quantity of tweets is already sufficient to identify distinct
events in time, that precede the election day, for instance, the
presentation of the political manifestos of the individual parties or
TV~shows covering the~election.

We start with the investigation of the influence of \emph{troll accounts} of
the Internet Research Agency~(IRA), which have been disclosed in the context
of the investigations of Russian interference in the 2016 US~presidential
elections~\citep{website:twitteriralist1, website:twitteriralist2}. We find
that \numtrollsfound of these trolls have also been active for the \btw,
resulting in a total amount of~\numtrolltweets~tweets in our~dataset.
Based on these first impressions we broaden our perspective to the entire
political landscape looking for indicators of propaganda.
In a detailed analysis, we survey specific topics and how these are related
to political parties as well as individual users that have contributed to
them.
For instance, topics related to the controversial right-wing party
\party{\mbox{Alternative für Deutschland}}~(AfD) have been predominant
during the election, including supporting as well as opposing positions.

Additionally, we develop a detector that is able to rate automated behavior
in order to identify \emph{bot~accounts}~in our dataset, which have been
identified for being a root cause for the amplification of
propaganda~\cite{WooHow16}.
Using this classifier we find \numtotalbots previously unknown bots, which
represent \perctotalbots of all user accounts in our dataset. While this
number seems surprisingly large, it is perfectly in line with previous
research, which states that \num{9}--\perc{15} of all active Twitter
accounts are~bots~\cite{VarFerDavMenFla17}.
However, differentiating the automated behavior of bots and the repetitive
manual actions of eagerly tweeting users is particularly~difficult. Thus our
results should be rather seen as first~indicators.

\clearpage

In summary, we make the following contributions:
\begin{itemize}%
    \item \textbf{Analysis of Known Actors.} We identify \numtrollsfound
    known actors involved in propaganda by correlating the published IRA
    troll accounts with the users from our~dataset.

    \item \textbf{Investigation of the Propaganda Landscape.} We analyze
    the largest dataset of tweets in the context of the \btw, in particular,
    \approxtotaltweets tweets over \numdays days, and inspect them
    regarding indicators of~propaganda.

    \item  \textbf{Detection of Automated Propaganda.} We effectively detect
    \numpredictbots~previously unknown bots that contribute to propaganda by
    implementing a classifier that can identify automated account behavior.

\end{itemize}

The remainder of the paper is organized as follows:
\sect{sec:dataset} discusses the basic properties of our dataset that
has been recorded during and prior to the \btw.
In \sect{sec:trolls}, we investigate the presence of known propaganda actors
in this data, before we discuss the overall political landscape of the
dataset regarding indicators of propaganda in
\sect{sec:propagandalandscape}.
Subsequently, we describe and evaluate our bot detector in
\sect{sec:bots}.
Related work is discussed in \sect{sec:relatedwork}, while
\sect{sec:conclusion} concludes the~paper.

\section{The German Federal Election on Twitter}
\label{sec:dataset}
For our analysis, we consider \approxtotaltweets tweets that have been
published in the context of the \btw~(\emph{Bundestagswahl}) and have
been collected over \numdays days, from \timespan.
As we are relying on the publicly available Twitter Stream, we receive
maximally \perc{1} of all publicly available tweets.
This limit, however, is only hit seldom. Due to random sampling, the
subsequently reported numbers can be safely extrapolated and the drawn
conclusions remain valid.
To restrict our analysis to the \btw, we apply the search~terms shown in
\tab{tab:searchterms}, that correspond to the abbreviations of the major
German parties\footnote{We consider all parties that have cleared the
\perc{5} threshold in the previous federal election (2013) or in one of the
previous state elections (2014 -- 2016). We additionally consider the NPD
that has closely failed the threshold (\perc{4.9}) in Saxony in 2014.}.
For \party{Die~Grünen} and \party{Die~Linke} we use different common
abbreviations, derived from the list of recognized parties by the Federal
Electoral~Committee~\cite{website:federalreturiningofficer}, as these do not
bear official~acronyms.

Based on a manual plausibility examination of the collected data on a sample
basis, we found an exceptionally high amount of tweets in Portuguese
language matching the search term \emph{fdp}.
Further investigation revealed that \emph{fdp} is a commonly used
abbreviation for a Portuguese swearword that is tainting our dataset.
Due to the fact that the language of the affected tweets is not correctly
identified by Twitter, we cannot use this feature for filtering.
Instead, we completely exclude all tweets that contain the search term
\emph{fdp}, which has accounted for \perctweetswithoutfdplost of the tweets.

In the following, we focus on the \numtweetswithoutfdp remaining tweets for
further~analysis.
We proceed with the detection of known propaganda actors in our~dataset.

\section{Known Actors} %
\label{sec:trolls}

In the course of the investigations of Russian interference in the 2016 US
presidential elections, Twitter has composed a list of accounts that are
linked to the Internet Research Agency (IRA)~\citep{website:twitteriralist1}
and had been
identified to be influential during the US~elections.
An updated list was forwarded to the US Congress in
June~2018~\citep{website:twitteriralist2} and released to the public to
foster further research on the behavior of those accounts~
\citep{website:schiffstatement}.

\begin{table}[t]
    \begin{minipage}{\columnwidth}
        \tablesize
        \centering
        \renewcommand*{\thempfootnote}{\fnsymbol{mpfootnote}}
        \caption{Search terms used for the data acquisition.}
        \label{tab:searchterms}
        \begin{tabularx}{\columnwidth}{p{7.6cm}p{4.8cm}@{\hskip 2mm}l}
            \toprule
            Party & Political Direction & Term\\
            \midrule
            \emph{Alternative für Deutschland (AfD)}
                 & Right-wing to far-right
                 & \texttt{afd}\\
            \emph{Christlich Demokratische Union (CDU)}
                 & Christ ian-democratic, liberal-conservative
                 & \texttt{cdu}\\
            \emph{Christlich-Soziale Union (CSU)}
                 & Christian-democratic, conservative
                 & \texttt{csu}\\
            \emph{\textcolor{gray}{Freie Demokratische Partei (FDP)}}
                 & \textcolor{gray}{(Classical) Liberal}
                 & \textcolor{gray}{\texttt{fdp}}\\
            \emph{Bündnis 90/Die Grünen}
                 & Green politics
                 & \texttt{gruene}\footnote{Additionally:
                                            \texttt{grüne},
                                            \texttt{diegruenen},
                                            \texttt{diegrünen}}\\
            \emph{Die Linke}
                 & Democratic socialist
                 & \texttt{linke}\footnote{Additionally:
                                           \texttt{dielinke}\vspace{-8mm}}\\
            \emph{Nationaldemokratische Partei Deutschlands (NPD)}
                 & Ultra-nationalists
                 & \texttt{npd}\\
            \emph{Sozialdemokratische Partei Deutschlands (SPD)}
                 & Social-democratic
                 & \texttt{spd}\\
            \bottomrule
        \end{tabularx}\\
    \end{minipage}
    \vspace{10mm}
\end{table}

Based on the assumption that existing Twitter accounts are often reused
for other purposes, we try to identify the same trolls in our dataset.
To this end, we match the list of the \numtrollsinlist published IRA
troll accounts to the user accounts from our dataset.
Since the screen name of a user account can be freely changed, we first map
the obtained screen names to their corresponding unique
user~IDs~\cite{ZanCauCriSirStrBla18}.
In doing so, we are able to detect \numtrollsfound of the IRA troll
accounts in our dataset which is \perctrollsfound of the total number of
users.
Surprisingly, only one of the identified accounts has changed its screen
name during this time.
However, the identified accounts are only responsible for a total
amount of \numtrolltweets~tweets that is \perctrolltweets of the tweets
from our dataset, rendering their potential direct influence comparably
low.
Interestingly, \perctrolltweetssmallerthanthirtyfive of the identified
accounts have tweeted less than \num{35} tweets over the entire time span,
while the top \num{3} troll accounts published more than
\num{1000} tweets each.
Similarly, to the entire dataset, most of the trolls' tweets are
actually retweets~(\numtrolltweetretweet); however, there is also a
significant amount of original tweets~(\numtrolltweetsorig) and fewer
quotes~(\numtrolltweetsquote).
Due to the fact that the list of IRA accounts was made publicly available a
significant time ago, it is likely that the IRA has created new accounts
that we are not aware of, yet.

\begin{figure}[htbp]
    \centering
    \subfloat[Creation dates of IRA troll accounts.]{%
        \label{fig:iratrollaccountsbydate}
        \includegraphics[width=0.45\columnwidth,trim=17pt 0pt 17pt 0pt,clip]
        {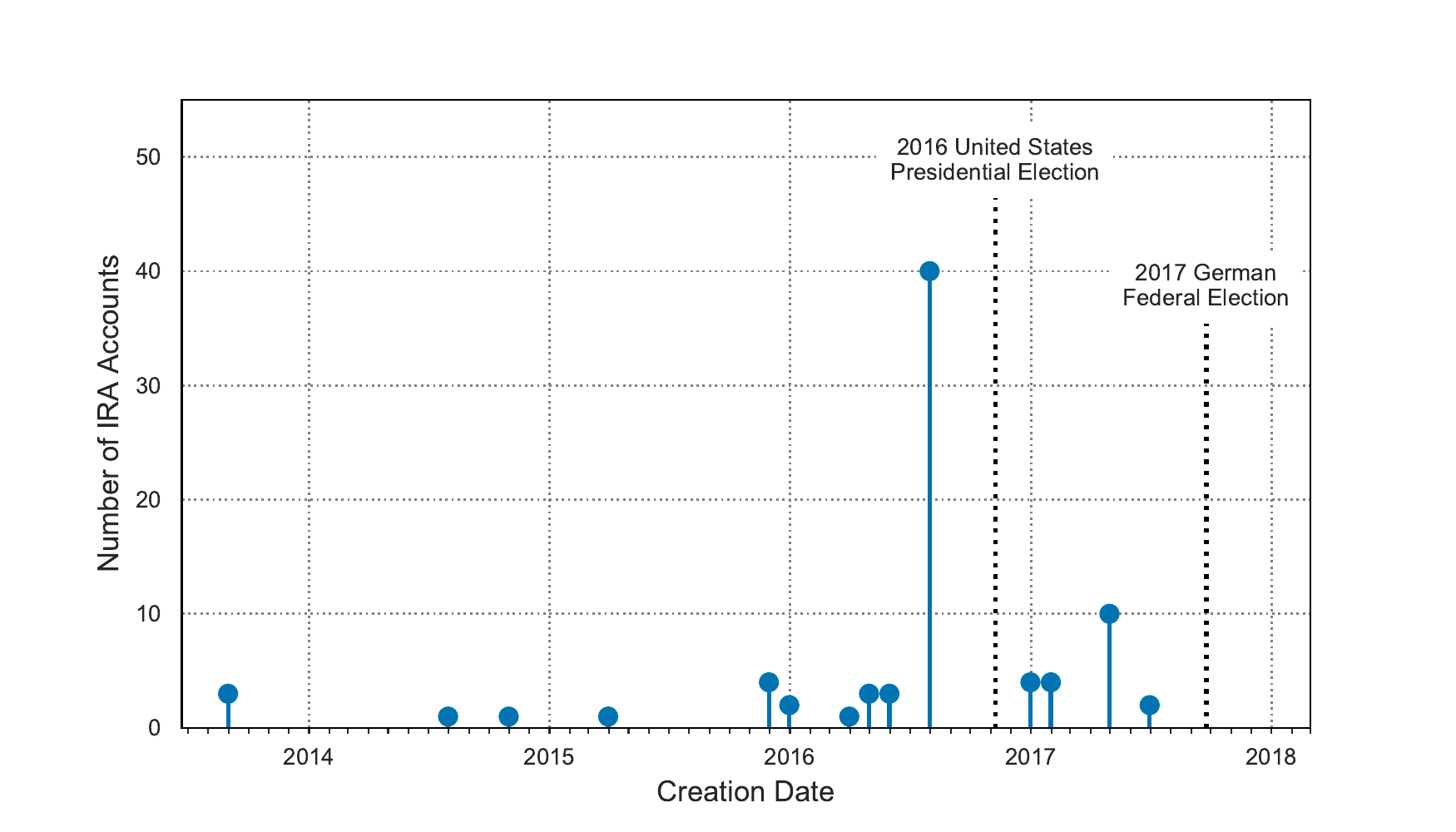}
    }
    \qquad
    \subfloat[Tweets posted in the context of the \btw.]{%
        \label{fig:iratrollaccountsbybtwcontribution}
        \includegraphics[width=0.45\columnwidth,trim=17pt 0pt 17pt 0pt,clip]
        {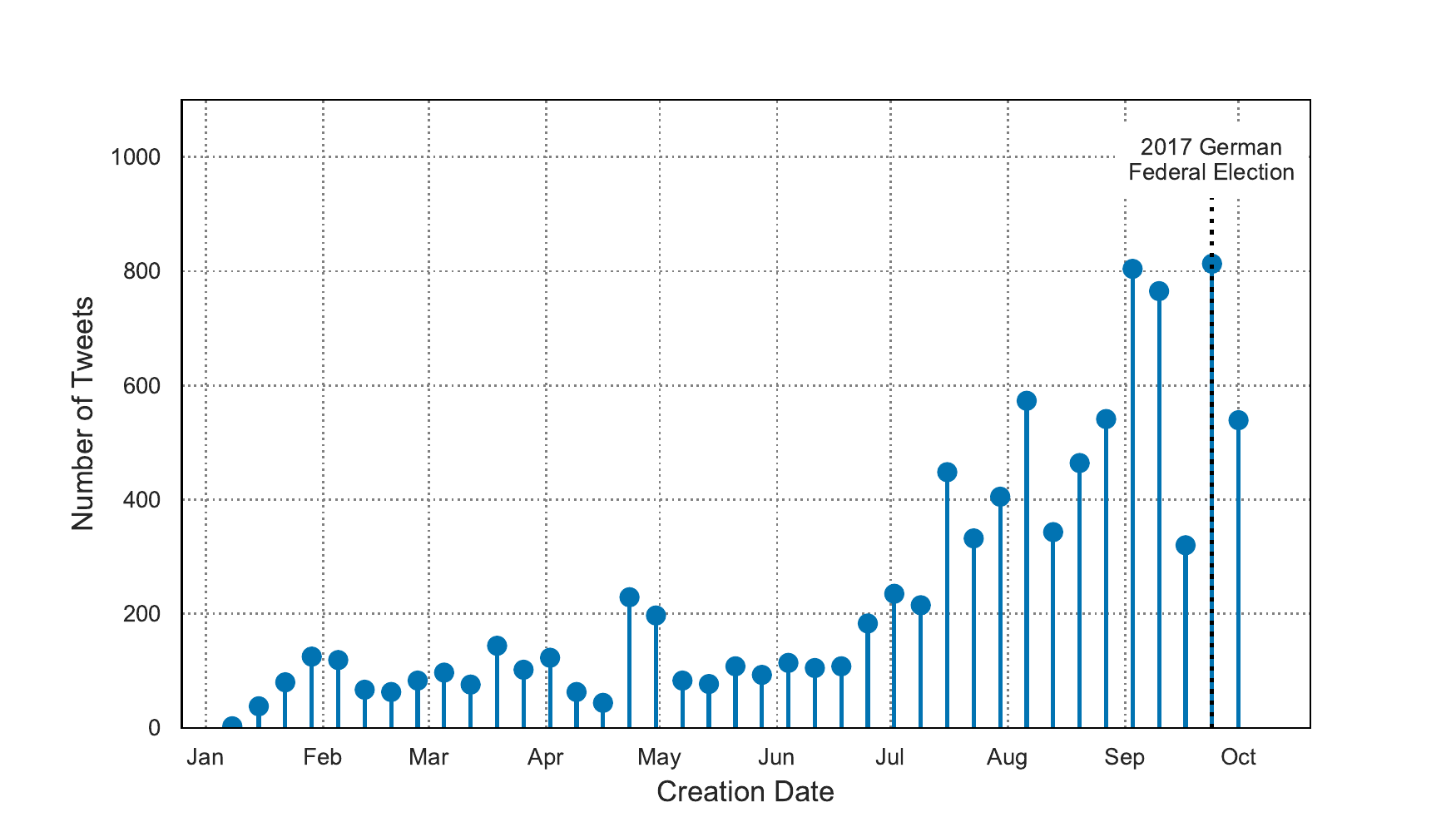}
    }
    \caption{Internet Research Agency (IRA) troll accounts.}
    \label{fig:iratrollaccounts}
\end{figure}

\fig{fig:iratrollaccountsbydate} shows the creation dates of the IRA
accounts over the last few years.
Most of the IRA accounts have been created before November 2016, the month
of the US presidential elections, with a significant peak in July 2016.
However, additional IRA accounts have been created between the beginning and
mid-2017 which means right before the \btw.
\fig{fig:iratrollaccountsbybtwcontribution} shows the number of tweet
contributions of the IRA accounts in the context of the 2017 \btw.
Unsurprisingly, there is a strong increase of tweets over the year 2017,
with its highest peaks at the beginning of September, the month of the
election, and particularly on the day of the election~itself.

Finally, to examine the impact of the IRA accounts on other users, we
verify if other accounts do interact with the IRA troll
accounts, for instance, by retweeting their tweets.
First of all \numtrollretweets of the tweets posted by IRA accounts
have been retweeted.
Only \numtrollretweetsbytrolls tweets originate from the known IRA accounts,
leaving the large remainder to other users.
Interestingly, the \numtrollquotes quoted tweets from IRA accounts have
all been quoted by other users, that are outside the peer group
of known IRA accounts.
Although the majority of the other users are likely regular user accounts,
there seem to be a fraction of accounts that are unknown troll accounts, we
are not aware of.
We conclude that although the amount of IRA accounts and corresponding
tweets is low, in comparison to the total amount of recorded users and
tweets, there is a verifiable impact from the IRA accounts on other
\mbox{accounts of the dataset.}

\section{Propaganda Landscape}
\label{sec:propagandalandscape}
Based on our analysis of known propaganda actors, we broaden our perspective
by taking the general political propaganda landscape into account.
To this end, we proceed with an analysis of the total tweet corpus to verify
if the same ratio of original tweets, retweets and quotes can be observed
for all collected tweets and parties.

\begin{figure*}[htbp]
    \centering
    \includegraphics[width=\textwidth,trim=0pt 0pt 0pt 0pt,clip]
    {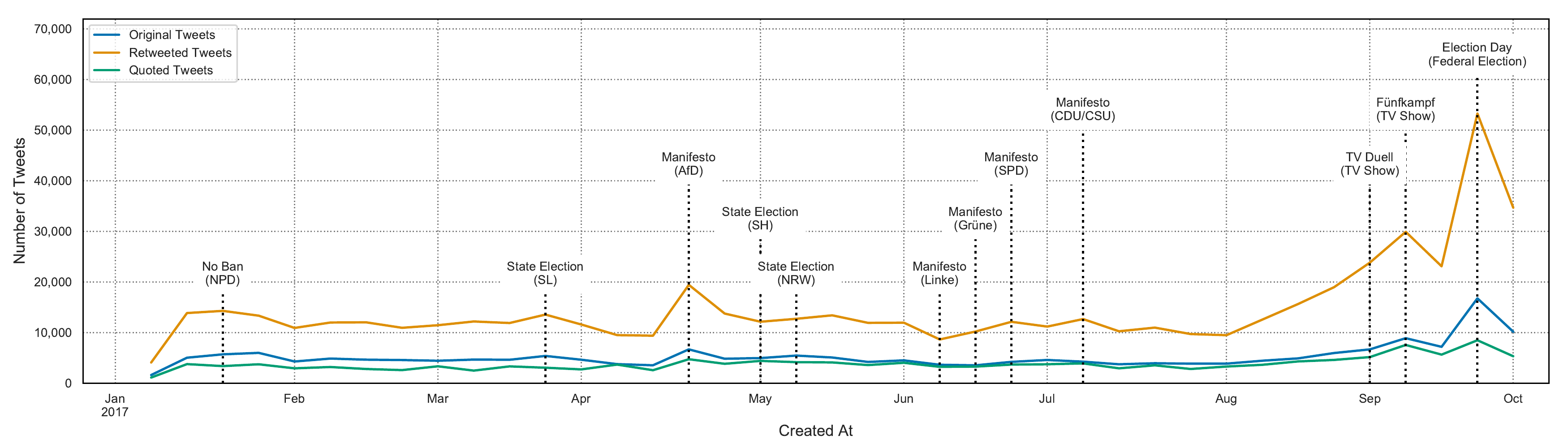}
    \caption{Development of tweet types over time.}
    \label{fig:tweettypes}
\end{figure*}

\fig{fig:tweettypes} shows the temporal development for original
tweets (blue), retweets (yellow), and quoted tweets (green).
Notice that the amount of retweets significantly exceeds the other two tweet
types. Consequently, these are a particularly strong factor of amplification
when spreading opinions.
Original or quoted tweets occur roughly \mbox{\num{60}$-$\perc{75}} less
frequent, each. However, the general trend leading up to the collection's
highest value at election day, and the shape of the amount's development
corresponds to all three types.

Throughout the recording, we observe local peaks that may be
attributed to distinct events in time, which we briefly discuss in the
following:
In~January the \emph{Federal Constitutional Court} has ruled in favor of not
banning the far-right, nationalists party \party{NPD}, which has been
preceded and succeeded by heated debates.
The state elections of \emph{Schleswig-Holstein}~(SH), \emph{Saarland}~(SL),
and \emph{North Rhine-Westphalia}~(NRW), in turn, have only triggered
mediocre response, whereas the presentation of the election manifestos for
the \btw partly receives significant attention.
Particularly, the publication of the manifesto of the right-wing party
\party{AfD} at the end of April is noteworthy at this point.
Starting in August, we record a strong increase of tweets leading up to
the federal election day on \nth{24} of September. This rise is
supported by several political talk shows, such as \emph{TV Duell} and
\emph{Fünfkampf} at the beginning~of~September.

To get a clearer view of the involved user accounts and topics, we further
analyze the most frequent hashtags, media files, and quoted/retweeted
user~accounts.

\subsubsection*{Hashtags}
\begin{figure}[tb]
    \centering
    \subfloat[Single hashtags]{%
        \label{fig:tophashtags}
        \includegraphics[width=0.45\columnwidth,trim=25pt 20pt 27pt 80pt,clip]
        {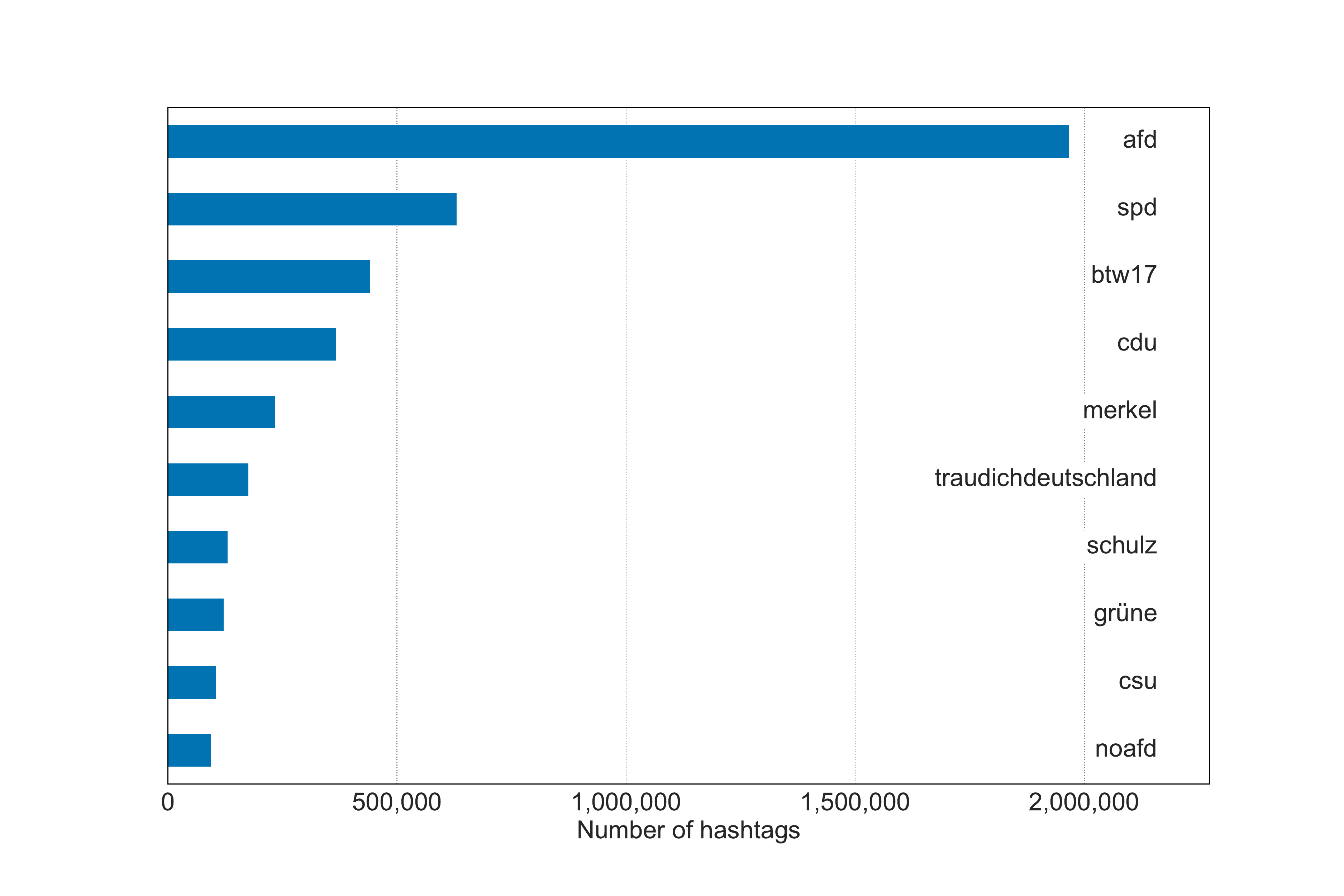}
    }
    \quad
    \subfloat[Combinations of hashtags]{%
        \label{fig:tophashtagcombinations}
        \includegraphics[width=0.45\columnwidth,trim=25pt 20pt 27pt 80pt,clip]
        {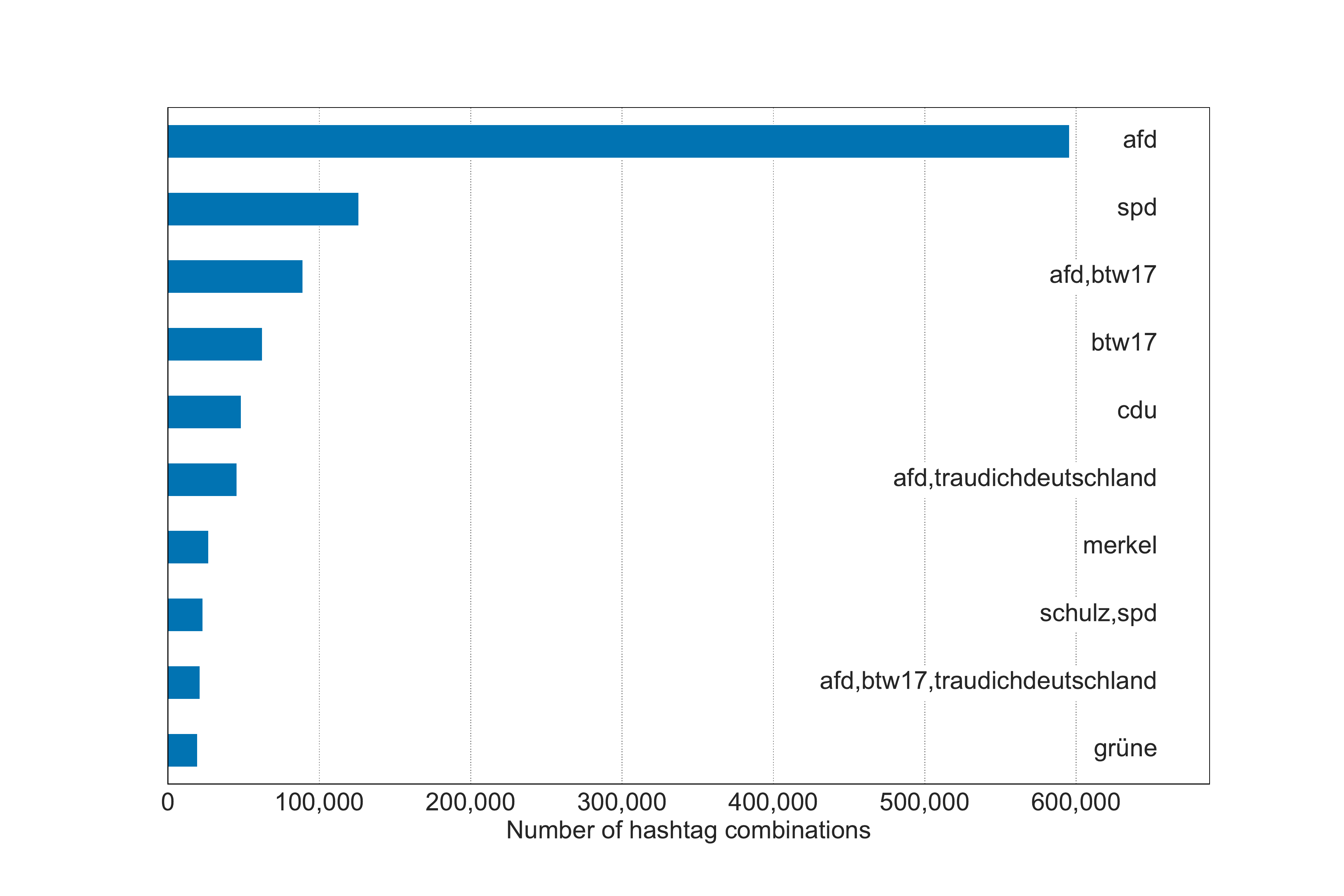}
    }
    \caption{Top-10 individual hashtags and combinations.}
\end{figure}

Among the ten most used hashtags we observe the acro\-nyms of five
political parties that have been up for election.
\fig{fig:tophashtags} shows a summary of the top \num{10} hashtags and
their number of occurrences.
Interestingly, the party that has triggered the largest peak in tweets when
presenting their election manifesto, the \party{AfD}, also peaks in total as
hashtag \hashtag{afd}, with \num{1968601} occurrences. Thereby, the
\party{AfD} occurs three times more often than the second-placed \party{SPD}
with \num{631209} occurrences.
The general hashtag for the \btw, \hashtag{btw}, in turn, is only used
in \num{442457} tweets.
On the sixth place, with the campaign \hashtag{traudichdeutschland}, the
\party{AfD} takes a prominent position for a second time with
\num{176677}~mentions.

Moreover, in \fig{fig:tophashtagcombinations} we consider the most used
combinations of hashtags and observe a similar dominance of the \party{AfD}.
The hashtag \hashtag{afd} appears in four out of ten different combinations.
In summary, the \party{Alternative für Deutschland (AfD)} seems to be
particularly active on Twitter in comparison to other parties.

\subsubsection*{Media}

\newlength{\topimageheight}
\setlength{\topimageheight}{3cm} %

\newlength{\topimagewidth}
\setlength{\topimagewidth}{5.4cm}

\begin{figure*}[tb]
    \centering
    \subfloat[Immigrant numbers in comparison to
              the location of the most \party{AfD} votes.]{%
        \label{fig:topmediaa}
        \includegraphics[width=\topimagewidth]
        {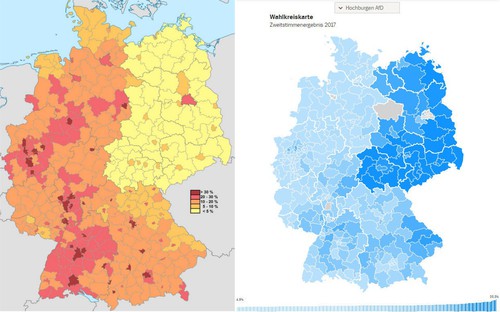}
    }
    \qquad
    \subfloat[Longish text about why eligible voters should
             not vote for \party{AfD}.]{%
       \label{fig:topmediab}
       \includegraphics[width=\topimagewidth, trim={0 12cm 0 0},clip]
       {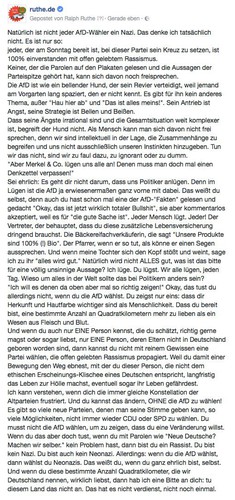}
    }
    \quad
    \subfloat[The former Chancellor \emph{Helmut~Kohl},
             (\cross 16. June, 2017).]{%
       \label{fig:topmediac}
        \includegraphics[height=\topimageheight,trim={5.5cm 0 5.5cm
        0},clip]
       {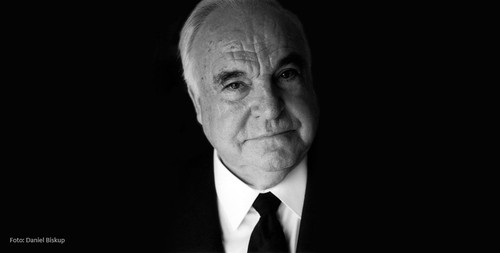}
    }
    \quad
    \subfloat[Author \emph{A. Moore} on protest votes.]{%
        \label{fig:topmediad}
        \includegraphics[height=\topimageheight]
        {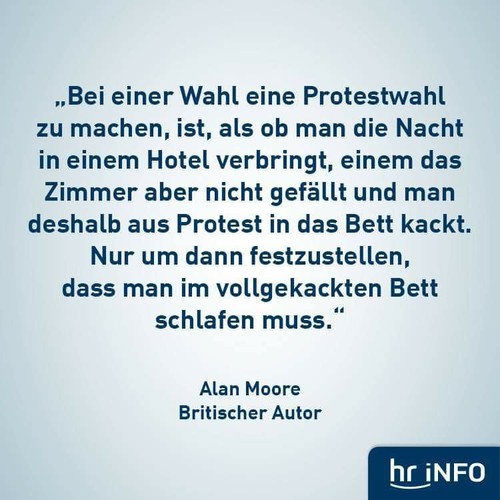}
    }
    \qquad
    \subfloat[Fake \party{AfD} election poster by \emph{heute-show}.]{%
        \label{fig:topmediae}
        \includegraphics[height=\topimageheight]
        {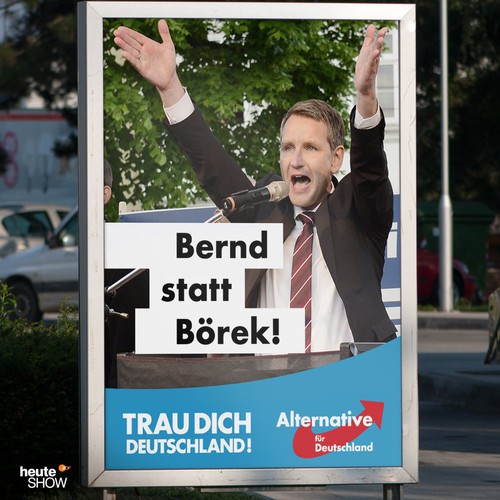}
    }
    \caption{Selection from most tweeted media from the dataset.}
    \label{fig:topfivemedia}
\end{figure*}

Next, we discuss the five most frequently tweeted images
that are related to the election (see \fig{fig:topfivemedia}).
With \numtopimagefirst occurrences, \fig{fig:topmediaa}, showing
two heat maps of Germany, is the most popular.
It displays the proportion of foreigners per region on the left and
the proportion of \party{AfD} voters per region on the right, showing a
drastic~imbalance.
The image in \fig{fig:topmediab}, tweeted \numtopimagesecond times, shows a
longish text about why eligible voters should not vote for the \party{AfD}.
Using an image for a long text was very common in the early days of Twitter,
since until November 2017 Twitter restricted the maximum number of
characters per tweet to \num{140}.
The third most tweeted picture shows a black and white portrait of the former
German Chancellor Helmut Kohl who died on the \nth{16} of June, 2017. This
news with the corresponding picture was tweeted \numtopimagethird~times.

The pictures shown in \fig{fig:topmediad} and \fig{fig:topmediae} occur
\numtopimagefourth and \numtopimagefifth times, respectively, and also
concern the \party{AfD}.
However, this images likewise popularize against the party by, on the one
hand, showing a comment of the British author A.~Moore explaining the idiocy
of protest votes and, on the other hand, displaying a fake \party{AfD}
election poster, that was published by the German political satire show
\emph{heute-show}.
Thus, the spike in hashtags likely cannot be traced back to the involvement
of supporters alone, but also to opponents of this controversial~party.

\subsubsection*{Quoted/Retweeted Users}

\begin{figure}[tbp]
    \centering
    \subfloat[Top \num{10} quoted users.]{%
        \label{fig:topquotedusers}
        \includegraphics[width=0.45\columnwidth,trim=25pt 20pt 27pt 80pt,clip]
        {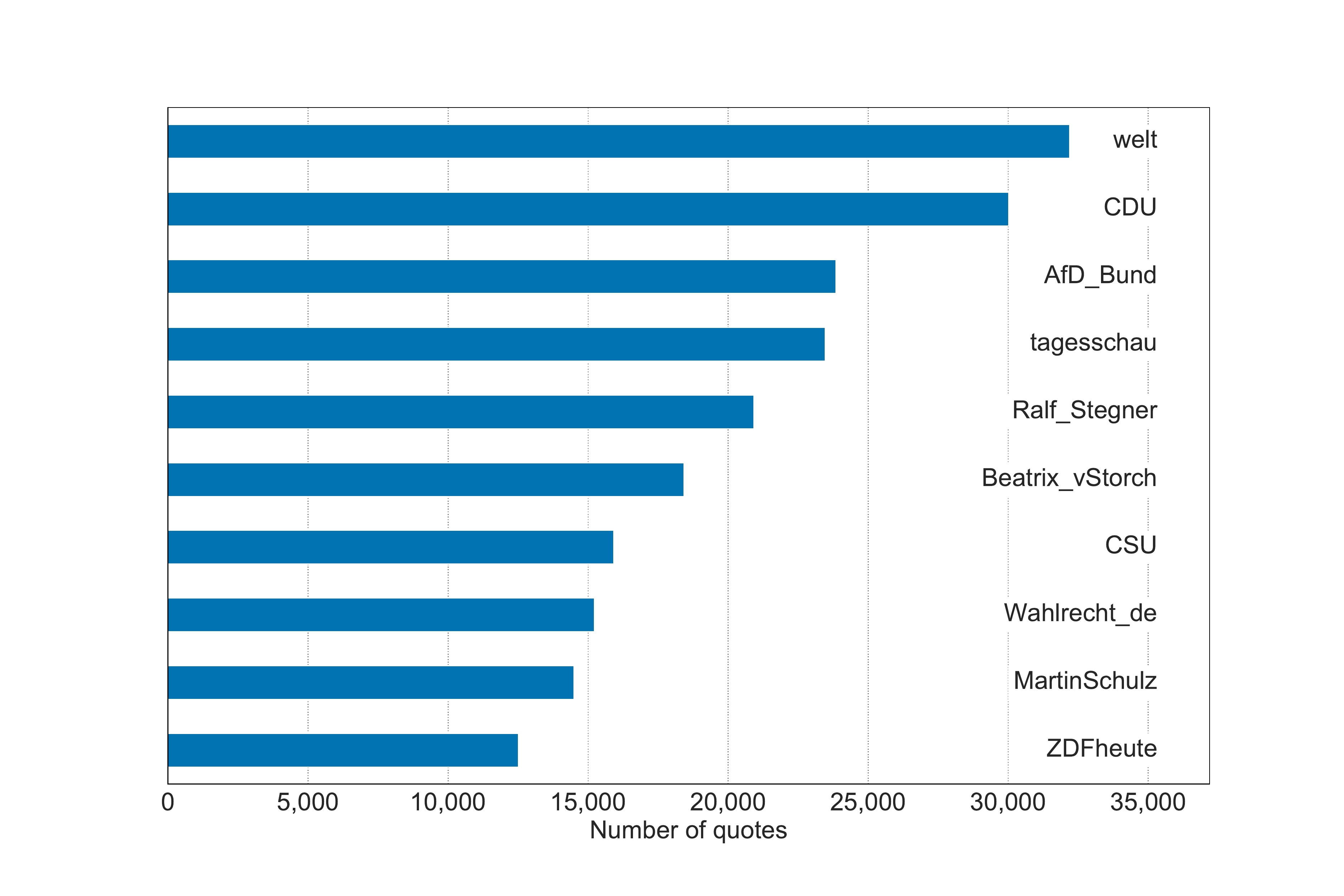}
    }
    \qquad
    \subfloat[Top \num{10} retweeted users.]{%
        \label{fig:topretweetedusers}
        \includegraphics[width=0.45\columnwidth,trim=25pt 20pt 27pt 80pt,clip]
        {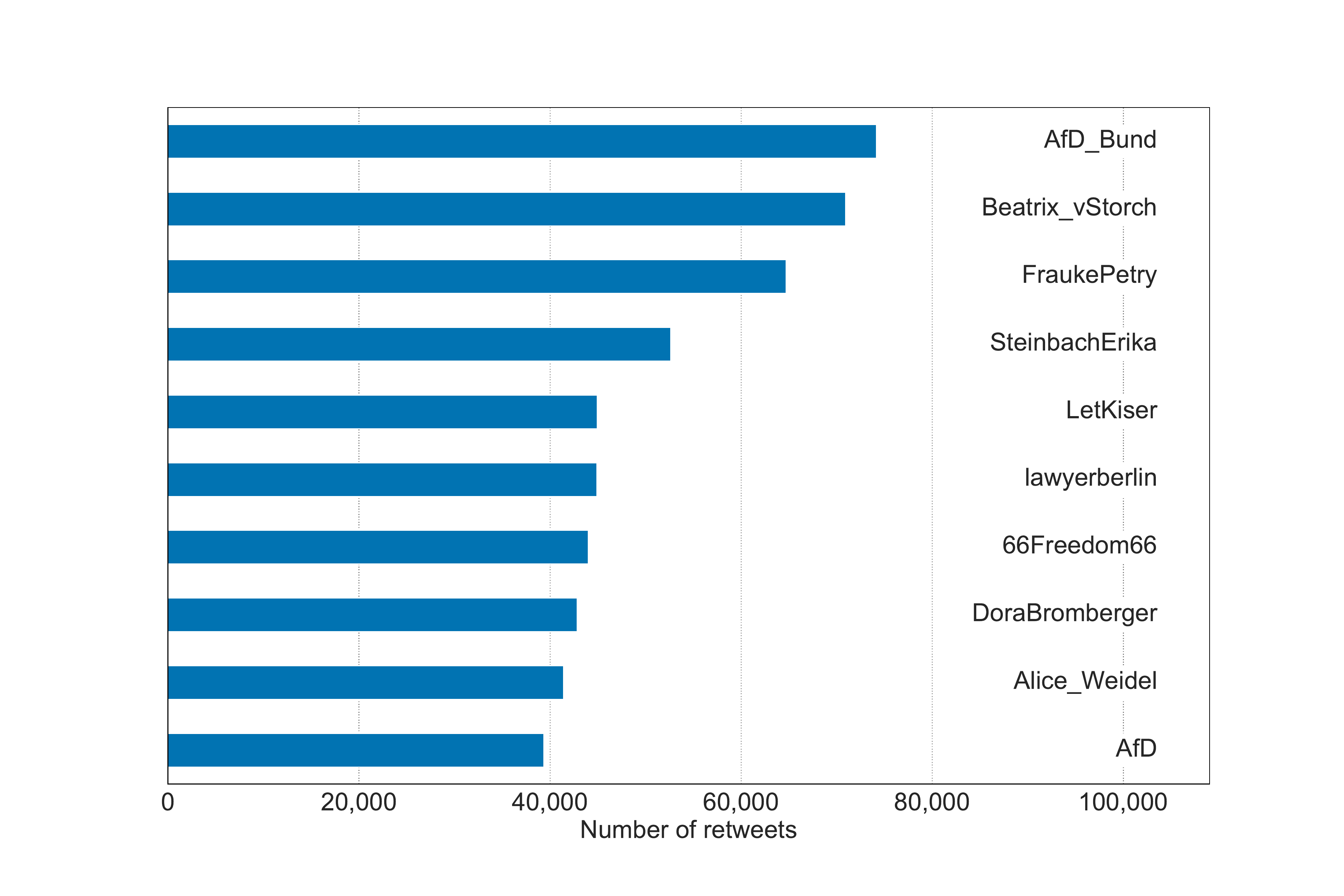}
    }
    \caption{Top \num{10} quotes and retweets.}
\end{figure}

As a measure of the popularity and influence of individual accounts, we
also look at the most quoted and retweeted users on our recording.
\fig{fig:topquotedusers} and \fig{fig:topretweetedusers} show the
top~\num{10} users for both categories.
Interestingly, \mention{AfD\_Bund}, and \mention{Beatrix\_vStorch} are
present in both rankings.
The first is the official account of the \party{AfD} party, and the latter
is an \party{AfD} politician, so are \mention{FraukePetry},
\mention{SteinbachErika}, \mention{lawyerberlin},
and\mention{Alice\_Weidel}.
Consequently, the list of the ten most retweeted users is largely dominated
by one party.
The remaining accounts, \mention{66Freedom66} and \mention{DoraBromberger},
advertise right-wing views and thus being also in line with the party.

Furthermore, three other political parties are rather prominently present:
\mention{CDU}, \mention{CSU}, and \mention{SPD}.
Especially the latter, the left-wing social democrats, have two politicians
among the top \num{10} quoted user accounts (\mention{Ralf\_Stegner} and
\mention{MartinSchulz}).
The remaining accounts mainly correspond to popular German news magazines:
\mention{welt}, \mention{tagesschau}, \mention{wahlrecht\_de}, and
\mention{ZDFheute}.

\section{Detecting Automated Propaganda}
\label{sec:bots}
Based on our findings on the political landscape in our dataset, we proceed
with the identification of automatic bot behavior, which holds responsible
for being one of the root causes for the amplification of heavily discussed
political topics~\cite{WooHow16}.
To this end, we apply a supervised machine learning approach to detect~bots.

Although the general topic of bot detection is well-known, the detection of
political social bots, in particular, is still an open challenge, as
indicated in related work~\cite[\eg][]{ChuGiaWanJaj12, FerVarDavMenFla16}.
On the one hand, this is due to its diverse characteristics, involving the
political direction and target audience, and, on the other hand, due to the
constant evolution of social bots that are approaching a more human-like
behavior by imitating common usage
patterns~\cite{FerVarDavMenFla16,QiAlBro18}.

For the implementation of our classifier, we make use of the insights gained
from the identified IRA trolls and saliences found in our in-depth analysis
of the political~landscape.

\subsubsection*{Labeling}
As the dataset has been just recorded for this purpose, there are no
existing labels of bots and humans, respectively, available that are
required to train a supervised machine learning model.
We therefore manually attribute Twitter accounts for both classes using a
set of simple heuristics.
These include a test for repetitive behavior of the same tweeting
pattern, a frequently posting of tweets without adhering sleep breaks at
least every~\SI{48}{\hour} or tweeting of multiple hashtags from the
trending topics combined with a URL, etc.
Even for trained experts, the distinction between humans and bot remains a
difficult challenge.
To avoid wrongly labeled training samples, we concentrate on those accounts
for which we could identify the class with high confidence.
As a result, we gathered \numtrainingbots bot and \numtraininghumans
human accounts in total for the training of the classifier.

\subsubsection*{Features}
Based on the heuristics that were used to manually label the training data,
we proceed with the engineering of additional features to improve the bot
detection rate by exploring the available tweet and user profile information
from our dataset.
We engineered \numfeatures unique features that are covering the four main
categories of metadata-based, text-based, time-based and user-based
features.
The \emph{metadata-based} features include features such as the average
number of tweets per day, the number of different clients used or the
retweet-to-tweet ratio.
In contrast, the \emph{text-based} features comprise, for instance, the
average tweet length, the vocabulary diversity or the URL ratio.
Furthermore, the \emph{time-based} features involve the longest average
break within \SI{48}{\hour} the median time between a retweet, the original
tweet, etc.
Finally, the \emph{user-based} features imply, for example, the number of
followers, the account verification status or the voluntary disclosure of
being a bot.
The complete list of derived features is presented in
\tab{tab:classifierfeatures}. %

\subsubsection*{Models}
We train and evaluate seven different machine learning algorithms for the
classification of bots and humans.
This includes the statistical-based \emph{LogisticRegression} model, the
non-parametric \emph{KNeighbors} model as well as the decision tree models
\emph{RandomForest}, \emph{AdaBoost} and \emph{GradientBoosting}.
Apart from that the two support vector machine learning variants
\emph{LinearSVC} and \emph{SVC} are applied and evaluated for their
aptitude.

We proceed with the application of our classifier in two experiments: a
controlled experiment and an extrapolation of our findings.
While the first controlled experiment targets the validation of our
classifier on the previously labeled training data and comparison to
Botometer~\cite{DavVarFerFlaMen16} as a baseline, the second extrapolate our
findings by applying the classifier on the remainder of our unlabeled
dataset as an indicator of the human-bot-ratio within the entire~dataset.

\subsection{Controlled Experiment}
Next, we apply the selected machine learning models to our training data by
making use of \mbox{\num{10}-fold} cross-validation and repeating the
experiments \numcrossvalrepetitions times, followed by averaging the result
metrics.
We identify the best parameter combination per classifier, by employing a
grid search, optimizing for the metric of best average \emph{Area Under
Curve~(AUC)}.
\tab{tab:testedclassifiers} shows the examined classifiers with the
best parameters found for each classifier type, sorted by the average AUC
overall repetitions in descending order.
We further compute the \emph{F1-Score} for a single value comparison that
considers both the precision and the recall likewise.
The best performance for each metric is shown in the table.
The best performing classifier, regarding the average AUC, is the
\mbox{\emph{GradientBoosting}} classifier with an AUC of \num{0.972} and
\num{0.1}-bounded AUC with~\num{0.907}.

\begin{table}[tbp]
    \caption{Results of the tested classifiers.}
    \label{tab:testedclassifiers}
        \fboxset{rounded,border-color=gray}
    \tablesize
    \centering
    \begin{tabularx}{1\columnwidth}{lXcc} %
        \toprule
        Classifier & ~
                            & Avg. F1-Score
                            & Avg. AUC\\
        \midrule
        GradientBoosting   &
                           & \lfbox{$0.891 \pm 0.033$}
                           & \lfbox{$0.976 \pm 0.011$}\\
        RandomForest       &
                           & $0.861 \pm 0.039$
                           & $0.972 \pm 0.013$\\
        AdaBoost           &
                           & $0.885 \pm 0.035$
                           & $0.971 \pm 0.013$\\
        SVC                &
                           & $0.851 \pm 0.038$
                           & $0.949 \pm 0.021$\\
        LogisticRegression &
                           & $0.841 \pm 0.043$
                           & $0.946 \pm 0.021$\\
        LinearSVC          &
                           & $0.821 \pm 0.040$
                           & $0.934 \pm 0.025$\\
        KNeighbors         &
                           & $0.704 \pm 0.060$
                           & $0.871 \pm 0.034$\\
        \bottomrule
    \end{tabularx}
\end{table}

\subsubsection*{Baseline}
\begin{figure}[tbp]
    \centering
    \subfloat[Full range]{%
        \includegraphics[width=0.45\columnwidth]{%
          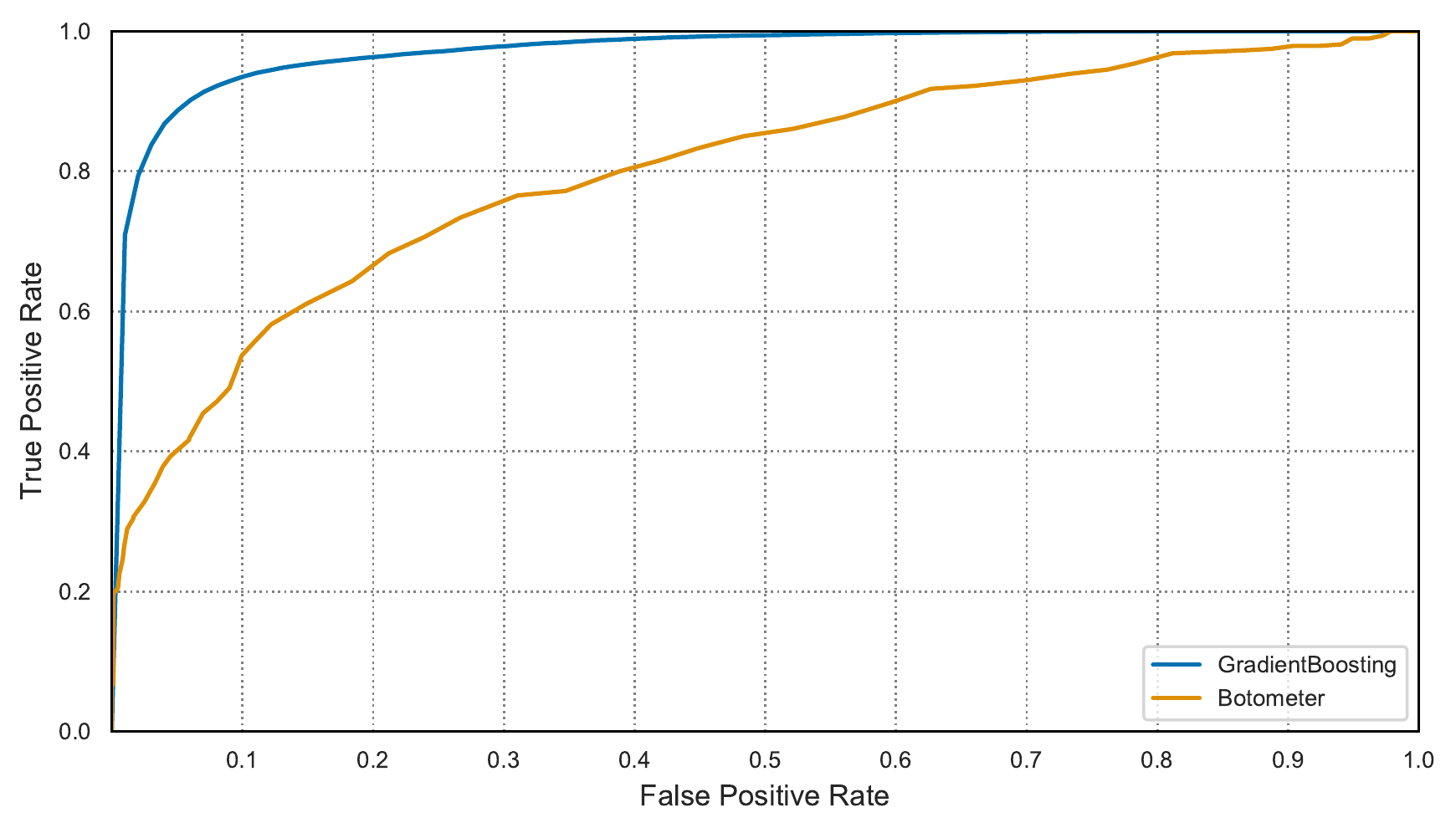}
        }
    \qquad
    \subfloat[False positives bound to \perc{10}]{%
        \includegraphics[width=0.45\columnwidth]
        {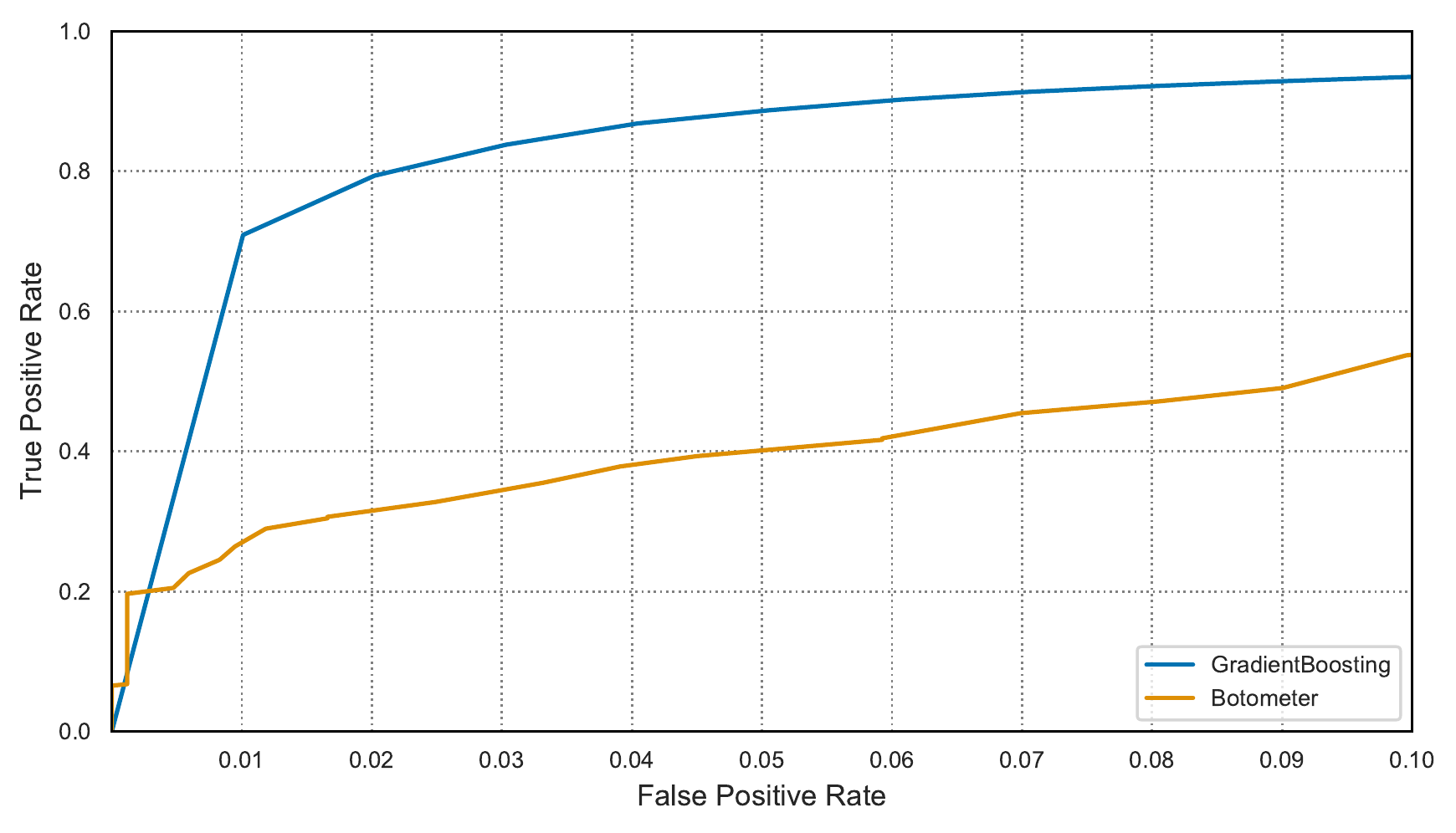}
        }
    \caption{Receiver Operating Characteristics (ROC) of
             GradientBoosting vs. Botometer.}
    \label{fig:gb_auc}
\end{figure}

As a baseline, we compare our results to the predictions of Botometer,
formerly known as BotOrNot~\cite{DavVarFerFlaMen16}, a popular bot
classifier that is publicly available on the Internet.
To this end, we query the Botometer~API for each of the previously labeled
Twitter accounts from the training dataset to obtain a corresponding bot
score.
The Botometer classifier yields an AUC of \num{0.802} and a value of
\num{0.679} if the false positive rate is bound to \num{0.1}, that is, a
false alarm rate of \perc{10}.
\fig{fig:gb_auc} shows the two ROC curves of Botometer and our improved
\emph{GradientBoosting} classifier. Our novel classifier outperforms the
mature Botometer classifier on our dataset by providing significantly
better results.

\subsection{Extrapolated Findings}
As an indicator of the human-bot-ratio within our entire dataset, we apply
the best performing classifier (\emph{GradientBoosting}) on the remainder of
our extracted user dataset. We focus on the \numinterestingusers potentially
interesting users that have published at least
\mbox{\numminimumnumberoftweets tweets} during the collection period.
Using our classifier we obtained predictions for
\mbox{\numpredicthumans human} and \mbox{\numpredictbots bot}
accounts.
In total, that means in combination with the previously manually labeled
accounts, we can identify \numtotalhumans human (\perctotalhumans) and
\numtotalbots bot (\perctotalbots) accounts within the potentially
interesting Twitter accounts.
Though we do not have labels for the complete user dataset to verify our
predictions, our results seem consistent with the recent study of
\citet{VarFerDavMenFla17}, who claim that between \perc{9} and
\perc{15} of active Twitter accounts are likely to be bots.

\section{Related Work}
\label{sec:relatedwork}
In the past, a plethora of research on various aspects of social media and
Twitter has been conducted.
In the following, we discuss the major points of contact with our
work:\\[-5mm]

\subsubsection*{Analyses of Political Elections.}
The first line of research deals with the analysis of political elections on
Twitter.
For instance, \citet{FraBelMer19} as well as \citet{PratSai2019} investigate
the 2015 and 2016 general elections in Spain.
While \citet{FraBelMer19} measure the regional support of political parties
on Twitter during the electoral periods in 2015 and 2016,
\citet{PratSai2019} focus on the two trending topics \hashtag{24M} and
\hashtag{Elections2015} on the election day in 2015 and build a predictive
model to infer the ideological orientation of~tweets.
Also the US 2016 presidential elections on Twitter are a topic of ongoing
research:
For instance, \citet{SaiYogNasSah19} characterize the Twitter networks of
the major presidential candidates, Donald Trump and Hillary Clinton, with
various American hate groups defined by the US Southern Poverty Law Center
(SPLC), while
\citet{CaeLimSanMar18} analyze the political homophily of users on Twitter
during the 2016 US presidential elections using sentiment analysis.

Furthermore, there are recent works on the 2017 German federal election:
\citet{GimHaaSchWit18} collect a representative dataset on the German
federal election and conduct a cluster analysis to derive eleven emergent
roles from the most active users, while
\citet{MorShaCalKar18} try to discover communities and their corresponding
themes during the \btw. Subsequently, they analyze how content is generated
by those communities and how the communities interact with each
other.\\[-5mm]

\subsubsection*{Bot Detection.} The second line of research deals with
the detection of bots on Twitter. Most recent works include
\citet{ChaHamMue16} who present a correlation finder to identify colluding
user accounts using \mbox{la-sensitive} hashing.
This has the advantage that no labels are required as for supervised
approaches.
In contrast, \citet{CrePiePetSpoTes17} study the phenomenon of social
spambots on Twitter and provide quantitative evidence for a paradigm-shift
in spambot design.
The authors claim that the new generation of bots imitates human behavior,
thus making them harder to~detect.

\citet{WalElo18} try to detect fake accounts that have been created by
humans. To this end, a corpus of human account profiles was enriched with
engineered features that had previously been used to detect fake accounts by
bots.
The tested supervised machine learning algorithms, could only detect the
fake accounts with a F1 score of \perc{49.75}, showing that human-created
fake accounts are much harder to detect than bot created accounts.

\citet{KudFer18} use a deep neural network based on contextual long
short-term memory (LSTM) to detect bots at tweet level. Using synthetic
minority oversampling, a large dataset is generated that is required to
train the model. As a result, an AUC of \num{0.99} is achieved.
Recently, \citet{CasAlPalAlfRamGonEloSan19} study the use of bots in the
2017 presidential elections in Chile. They manually derive labels for the
training data and then build a classifier for detecting bots. Though the
model reached good results in the training stage, the testing results were
not as good as they hoped.

\clearpage

In comparison to the above-mentioned classifiers, our detector makes use of
features from multiple categories of different domains  \ie metadata, text,
time and user-profile, to cover all aspects of modern bot behavior.

\section{Conclusion}
\label{sec:conclusion}

We have analyzed a total of \approxtotaltweets~tweets to investigated the
dissemination of propaganda in the context of the \btw.
We find that \numtrollsfound~of the trolls of Internet Research
Agency~(IRA) that have already been influencing the US~presidential
elections in 2016 have also been active a year later in Germany.

Based on these finding and the knowledge about the significance of
retweets and quoted tweets for propaganda purposes, we have then
broadened our analysis to the general political landscape. In this
scope, we have particularly inspected the most tweeted hashtags and
images as well as the involved users.
Our evaluation shows that especially the right-wing party \party{AfD}
has played a prominent role in several controversial discussions.
The hashtag \hashtag{afd}, for instance, dominates the top-10 ranking of
hashtag combinations and also the most retweeted users are \emph{all}
involved with this right-wing party.
Given the partly significant influence on the public discourse on
Twitter, it remains an open question whether this influence is driven by
automated efforts and bots.
The detector we have developed has enabled us to identify
\numtotalbots~previously unknown bots in our dataset, which account for
\perctotalbots~of all user~accounts.

The large proportion of automated accounts highlights the potential
danger when used for propaganda purposes.
While it has been inconclusive whether the propaganda efforts observed
in our dataset is mainly attributable to bot accounts, our study of the
\btw clearly shows that the political landscape heavily relies on
propaganda on social media. Particularly troublesome is the amount of
right-wing positions featured in the~data.

\clearpage

\bibliographystyle{ACM-Reference-Format}
\bibliography{bib/other}

\appendix
\section*{}
\label{sec:appendix}

\begin{table}[htbp]
    \scalebox{0.8}{%
    \begin{minipage}[t]{\textwidth}
        \caption{Engineered features used for classifying automated
                 propaganda, subdivided by four categories.}
        \label{tab:classifierfeatures}
            \centering
    \begin{tabularx}{\textwidth}{p{0.5cm}l@{\hspace{2em}}X}
        \toprule
            & Feature & Description\\

        \midrule
        \multirow{14}{*}{\rot{Metadata-based}}
            & \code{avg\_tweets\_per\_day} & Average number
                                             of tweets per day.\\
            & \code{total\_tweets} & Total number of tweets.\\
            & \code{orig\_ratio} & Ratio of own composed tweets to
                                   total tweets\\
            & \code{retweet\_ratio} & Ratio of retweeted tweets to
                                      total tweets.\\
            & \code{quote\_ratio} & Ratio of quoted tweets to total
                                    number of tweets.\\
            & \code{reply\_ratio} & Ratio of replies to total number
                                    of tweets.\\
            & \code{twitter\_client} & Used Twitter client
                                       ($\rightarrow$ terms mapped
                                        via tf-idf).\\
            & \code{official\_client} & Use of the official Twitter
                                        client.\\
            & \code{total\_clients} & Total number of used Twitter
                                      clients.\\
            & \code{unique\_users\_retweet\_ratio} & Ratio of unique
                                                     users in
                                                     retweets.\\
            & \code{unique\_users\_quotes\_ratio} & Ratio of unique
                                                    users in quotes.\\
            & \code{unique\_users\_retweet\_ratio} & Ratio of unique
                                                     users in
                                                     replies.\\
            & \code{longest\_conversation} & Longest conversation
                                             with a user.\\
            & \code{unique\_users\_conv\_ratio} & Ratio of unique
                                                  users in
                                                  conversations.\\

        \midrule
        \multirow{17}{*}{\rot{Text-based}}
            & \code{avg\_text\_len} & Average length of tweet text.\\
            & \code{std\_text\_len} & Standard Deviation of the length
                                      of tweet text.\\
            & \code{url\_ratio} & Ratio of tweets with URL.\\
            & \code{unique\_url\_ratio} & Ratio of unique URLs in
                                          tweets.\\
            & \code{unique\_url\_host\_ratio} & Ratio of unique
                                                host names in URLs.\\
            & \code{vocabulary\_diversity} & Diversity of used
                                             vocabulary in tweets.\\
            & \code{mentions\_ratio} & Ratio of mentions to tweets.\\
            & \code{hashtags\_ratio} & Ratio of hashtags to tweets.\\
            & \code{unique\_mentions\_ratio} & Ratio of unique mentions
                                               to tweets.\\
            & \code{unique\_hashtags\_ratio} & Ratio of unique hashtags
                                               to tweets.\\
            & \code{ending\_hashtags\_ratio} & Ratio of tweets that
                                               ends with a hashtags.\\
            & \code{starting\_mention\_ratio} & Ratio of tweets that
                                                starts with a
                                                mention.\\
            & \code{starting\_rt\_ratio} & Ratio of tweets that
                                           starts with RT.\\
            & \code{zip\_ratio} & Ratio of tweets after zipping
                                  to original size.\\
            & \code{user\_simhash} & Simhash of all tweets per user.\\
            & \code{avg\_duplicate\_simhash} & Average of tweets that
                                               have similar simhash.\\
            & \code{duplicate\_simhash\_ratio} & Ratio of all
                                                 duplicates to
                                                 amount of
                                                 from-users.\\

        \midrule
        \multirow{5}{*}{\rot{Time-based}}
            & \code{chi\_square\_seconds} & $\chi^2$-distribution of
                                            seconds of tweet
                                            creations.\\
            & \code{avg\_longest\_break} & Longest break of a user
                                           every \SI{48}{\hour} on avg.\\
            & \code{avg\_second\_longest\_break} & Second longest break
                                                   of a user every
                                                   \SI{48}{\hour} on avg.\\
            & \code{median\_retweet} & Median timespan between
                                       retweet \& orig. tweet.\\
            & \code{median\_quote} & Median timespan between a
                                     quote \& orig. tweet.\\

        \midrule
        \multirow{8}{*}{\rot{User-based}}
            & \code{total\_friends} & Total number of friends.\\
            & \code{total\_followers} & Total number of followers.\\
            & \code{friend\_follower\_ratio} & Ratio of number of
                                               friends to number of
                                               followers.\\
            & \code{has\_default\_profile\_image} & Has the default
                                                    profile image.\\
            & \code{has\_default\_user\_image} & Has the default user
                                                 image.\\
            & \code{is\_verified} & Is a verified Twitter account.\\
            & \code{has\_geo\_coordinates} & User has geo
                                             coordinates enabled.\\
            & \code{self\_bot} & User account contains the term `bot'
                                 in name.\\
        \bottomrule
    \end{tabularx}
        \end{minipage}
    }
\end{table}

\begin{backpage}
    \showtubslogo
    \begin{titlerow}[bgcolor=\clrDark,fgcolor=tubsWhite]{3}
        \large
          Technische Universität Braunschweig\\
          Institute of System Security\\
          Rebenring 56\\
          38106 Braunschweig\\
          Germany
    \end{titlerow}

    \begin{titlerow}[bgcolor=\clrLight]{4}
    \large
    ~
    \end{titlerow}
\end{backpage}

\end{document}